%% file: aaai2026.tex
\documentclass[letterpaper]{article} 
\usepackage{aaai2026}  
\usepackage{times}  
\usepackage{helvet}  
\usepackage{courier}  
\usepackage[hyphens]{url}  
\usepackage{graphicx} 
\usepackage{natbib}  
\usepackage{caption} 
\usepackage{algorithm}
\usepackage{algorithmic}

\usepackage{amsfonts}       
\usepackage{amssymb}
\usepackage{mathtools}
\usepackage{amsthm}
\usepackage{physics}
\usepackage{nicefrac}       
\usepackage{microtype}      
\usepackage[dvipsnames]{xcolor}
\usepackage[capitalize,noabbrev]{cleveref}

\usepackage{relsize}
\usepackage{graphicx}       
\usepackage{booktabs}       
\usepackage{multirow}       
\usepackage{subcaption}
\usepackage{colortbl}
\usepackage{tablefootnote}
\usepackage{algorithmic,algorithm}

\usepackage{color}

\usepackage{tabularx}
\usepackage{arydshln}
\usepackage{gensymb}
\usepackage{subfig}
\usepackage{makecell}
\usepackage[justification=raggedright, singlelinecheck=false]{caption}
\usepackage{pifont}
\usepackage{amsmath}
\usepackage{bm}

\usepackage{newfloat}
\usepackage{listings}
\DeclareCaptionStyle{ruled}{labelfont=normalfont,labelsep=colon,strut=off} 
\floatstyle{ruled}
\newfloat{listing}{tb}{lst}{}
\floatname{listing}{Listing}
\title{Lightning Fast Caching-based Parallel Denoising Prediction for Accelerating Talking Head Generation}
\author {
    Jianzhi Long\textsuperscript{\rm 1},
    Wenhao Sun\textsuperscript{\rm 2},
    Rong-Cheng Tu\textsuperscript{\rm 2}$^*$,
    Dacheng Tao\textsuperscript{\rm 2}$^*$
}
\affiliations {
    \textsuperscript{\rm 1}The University of Sydney\\
    \textsuperscript{\rm 2}Nanyang Technological University\\
    jlon7198@uni.sydney.edu.au, \{wenhao006, rongcheng.tu, dacheng.tao\}@ntu.edu.sg \\
}
\makeatletter
\def\showauthors@on{T}   
\makeatother

\usepackage{bibentry}

\begin{document}

\maketitle

\maketitle
\renewcommand{\thefootnote}{\fnsymbol{footnote}}
\footnotetext[1]{Corresponding authors.}

\begin{abstract}
Diffusion-based talking head models generate high-quality, photorealistic videos but suffer from slow inference, limiting practical applications. 
Existing acceleration methods for general diffusion models fail to exploit the temporal and spatial redundancies unique to talking head generation. 
In this paper, we propose a task-specific framework addressing these inefficiencies through two key innovations. 
First, we introduce \textbf{Lightning-fast Caching-based Parallel denoising prediction (LightningCP)}, caching static features to bypass most model layers in inference time.
We also enable parallel prediction using cached features and estimated noisy latents as inputs, efficiently bypassing sequential sampling.
Second, we propose \textbf{Decoupled Foreground Attention (DFA)} to further accelerate attention computations, exploiting the spatial decoupling in talking head videos to restrict attention to dynamic foreground regions.
Additionally, we remove reference features in certain layers to bring extra speedup.
Extensive experiments demonstrate that our framework significantly improves inference speed while preserving video quality.
\end{abstract}


\input{introduction}

\input{related_works}

\input{preliminaries}

\input{methods}

\input{experiments}

\input{conclusion}

\section{Ethical Statement}


This research is intended to support the development of ethical virtual humans and creative applications in entertainment and communication. 
However, as with other generative talking head technologies, there is a risk that the proposed acceleration framework could be misused for malicious purposes such as fraud, impersonation, or defamation \cite{liu2025t2v}. 
Moreover, the increasing prevalence of AI-generated media may contaminate public datasets and negatively impact the training of downstream AI models \cite{tu2025prospective, tu2025global}, potentially leading to degraded or unsafe system behavior.

To mitigate these risks, we recommend that any generated talking head videos be clearly watermarked or labeled as AI generated. 
We are also committed to assist the ongoing community efforts to develop robust evaluations and safety protocols \cite{tu2025automatic}, for promoting safe and trustworthy deployment of generative technologies.

\section{Acknowledgements}
This research is supported by the RIE2025 Industry Alignment Fund – Industry Collaboration Projects (IAF-ICP) (Award I2301E0026), administered by A*STAR, as well as supported by Alibaba Group and NTU Singapore through Alibaba-NTU Global e-Sustainability CorpLab (ANGEL).

\bibliography{aaai2026}



\end{document}

%% file: introduction.tex
\section{Introduction}

Talking head generation has received significant attention due to its wide applications in virtual avatars, digital content creation, and real-time communication \cite{prajwal2020lip, zhang2023sadtalker, wang2022one, tian2024emo}. 
Recent advancements in generative diffusion models \cite{rombach2022high, tu2024spagent} have greatly enhanced the realism and quality of talking head videos \cite{tian2024emo, xu2024hallo, chen2025echomimic, zheng2024memo}, surpassing earlier GAN-based techniques \cite{wang2021audio2head, zhang2023sadtalker, wang2022one}.  
However, the computational cost associated with diffusion models—primarily due to iterative denoising steps and large parameter sizes—results in slow inference and hinders real-time deployment \cite{sun2025vorta, sun2024asymrnr}. 
To accelerate diffusion models, previous methods employ model pruning \cite{fang2023structural} and distillation \cite{salimans2022progressive, meng2023distillation} techniques, but these often require expensive training.


\begin{figure}[h]
  \centering
  \begin{subfigure}{0.50\columnwidth}
    \includegraphics[width=\linewidth]{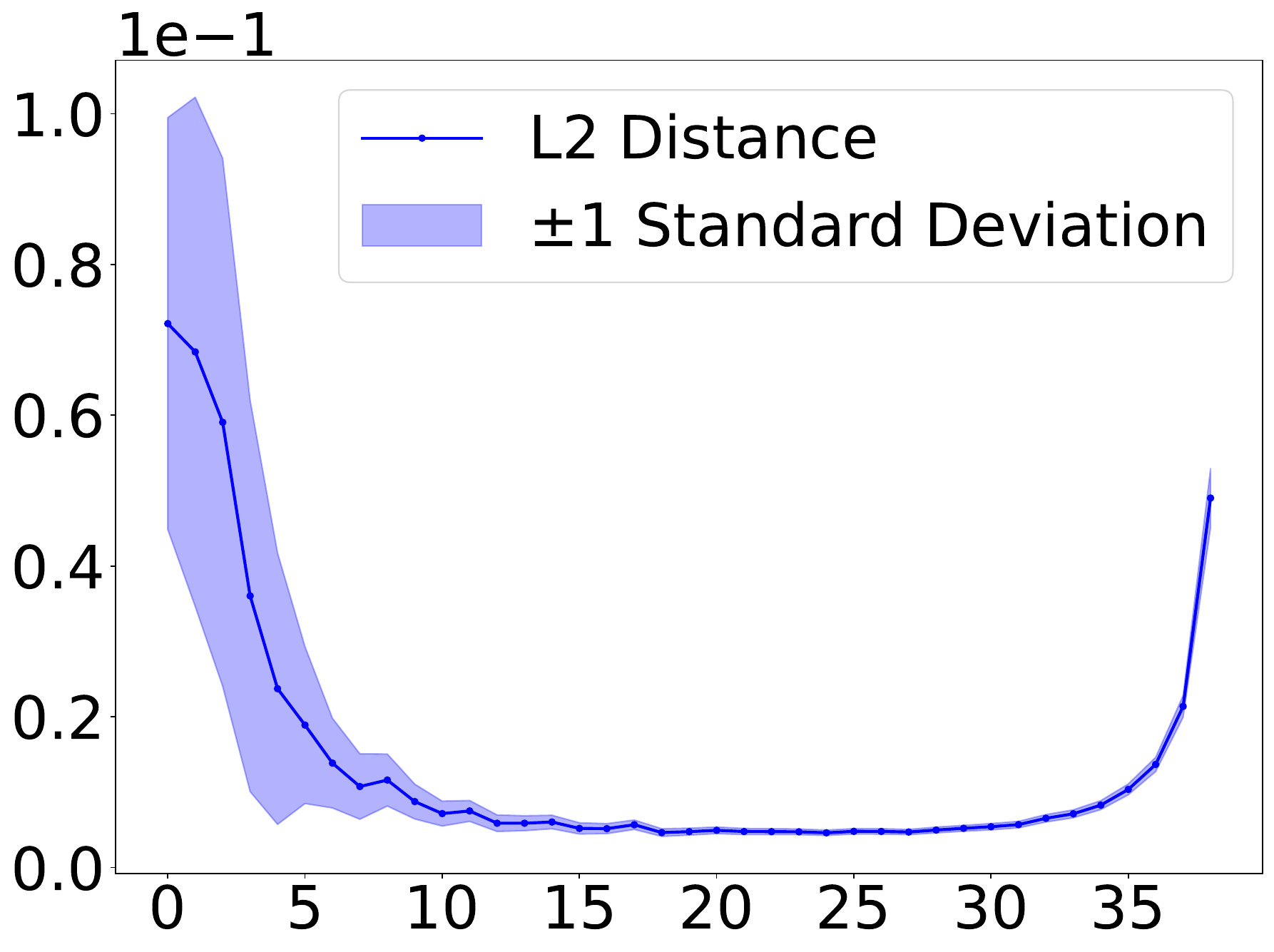}
    \caption{$L_2$ Distance.}
    \label{f_U31_mse}
  \end{subfigure}
  \hfill
  \begin{subfigure}{0.42\columnwidth}
    \includegraphics[width=\linewidth]{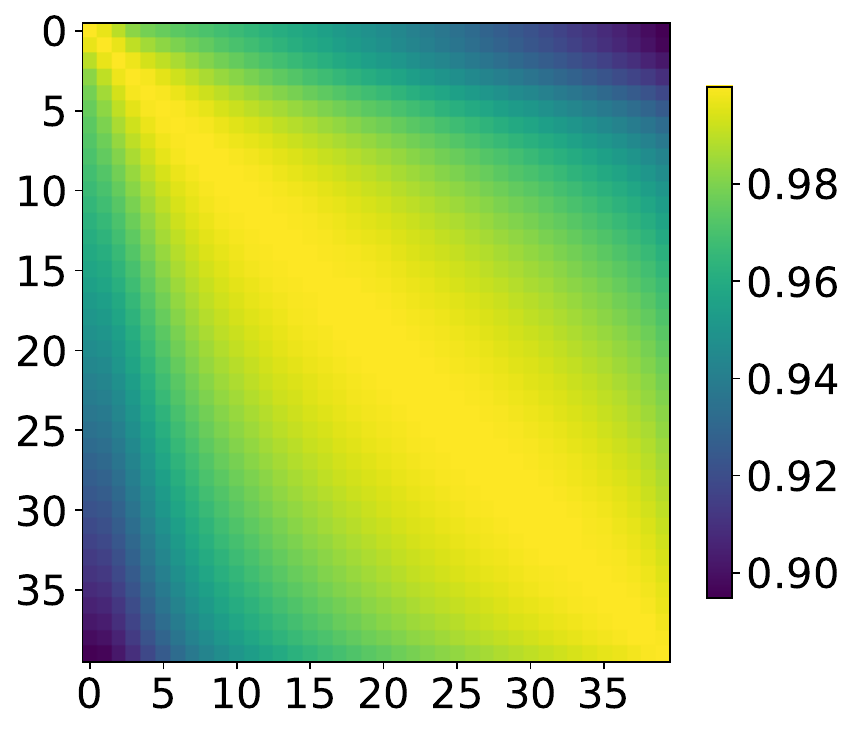}
    \caption{Cosine similarity matrix.}
    \label{f_U31_cosine_similarity}
  \end{subfigure}
  \caption{
  Analysis of the feature $f_{U_{31}}$ across timesteps in the Hallo model: (a) The $L_2$ Distance of the feature $f_{U_{31}}$ between consecutive timesteps, (b) The cosine similarity matrix of the feature $f_{U_{31}}$ between all timesteps.
  }
  \label{fig:feature_temporal_analysis}
\end{figure}


\begin{figure*}[h]
  \centering
  \begin{subfigure}{0.19\linewidth}
    \includegraphics[width=\linewidth]{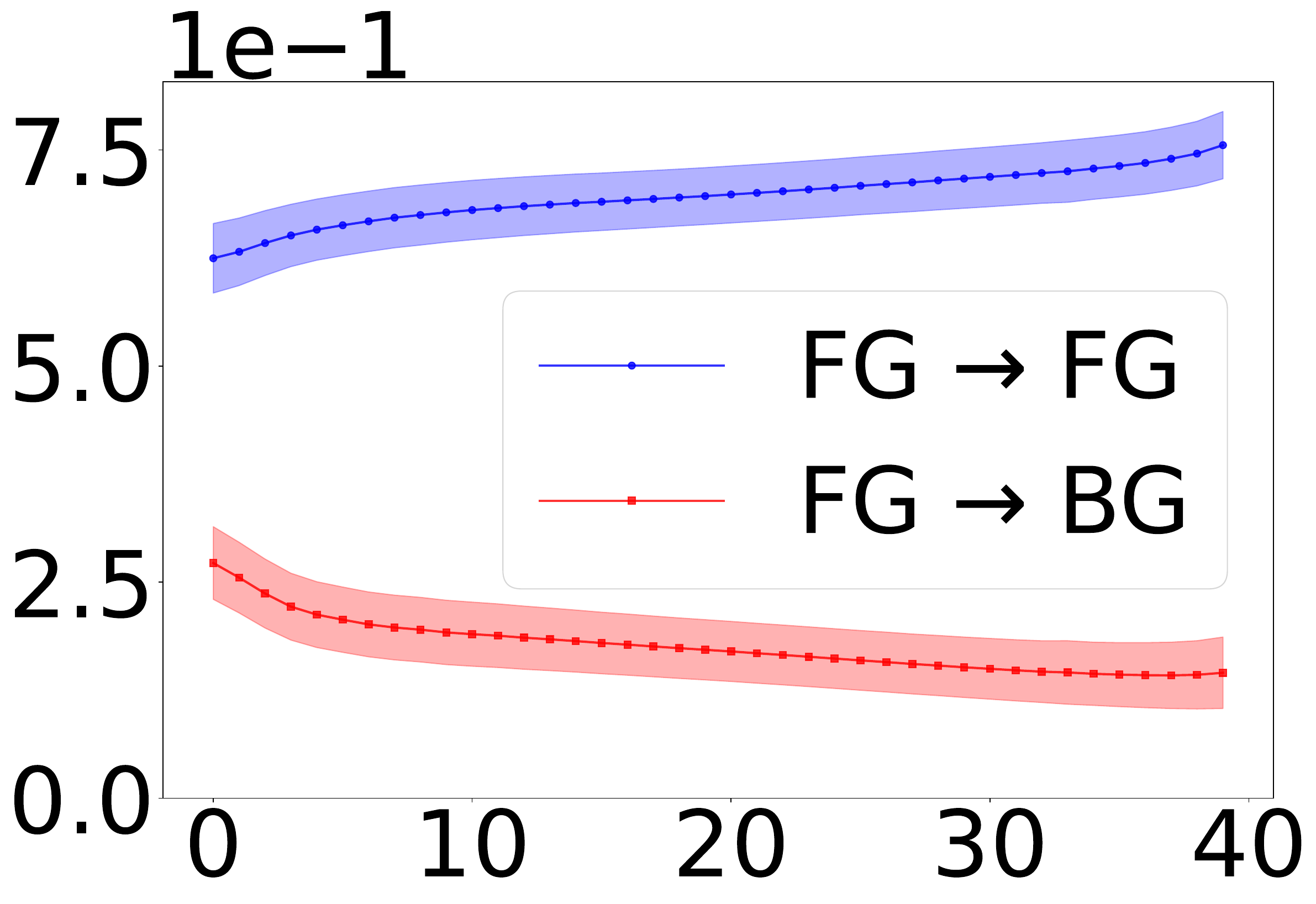}
    \caption{FG noisy latent feature.}
    \label{fig:fg_attention_weight_noise}
  \end{subfigure}
  \begin{subfigure}{0.19\linewidth}
    \includegraphics[width=\linewidth]{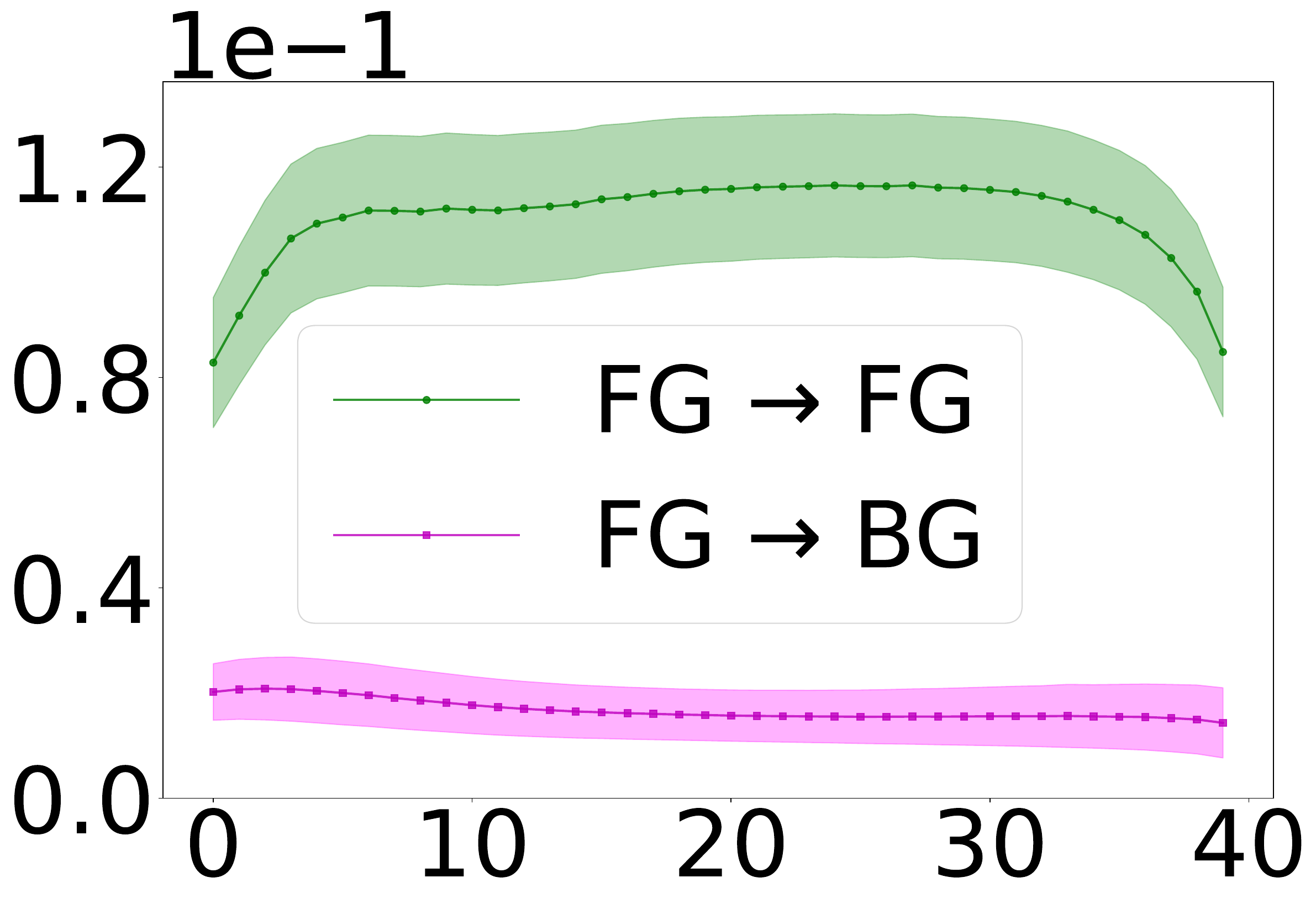}
    \caption{FG reference feature.}
    \label{fig:fg_attention_weight_reference}
  \end{subfigure}
  \begin{subfigure}{0.19\linewidth}
    \includegraphics[width=\linewidth]{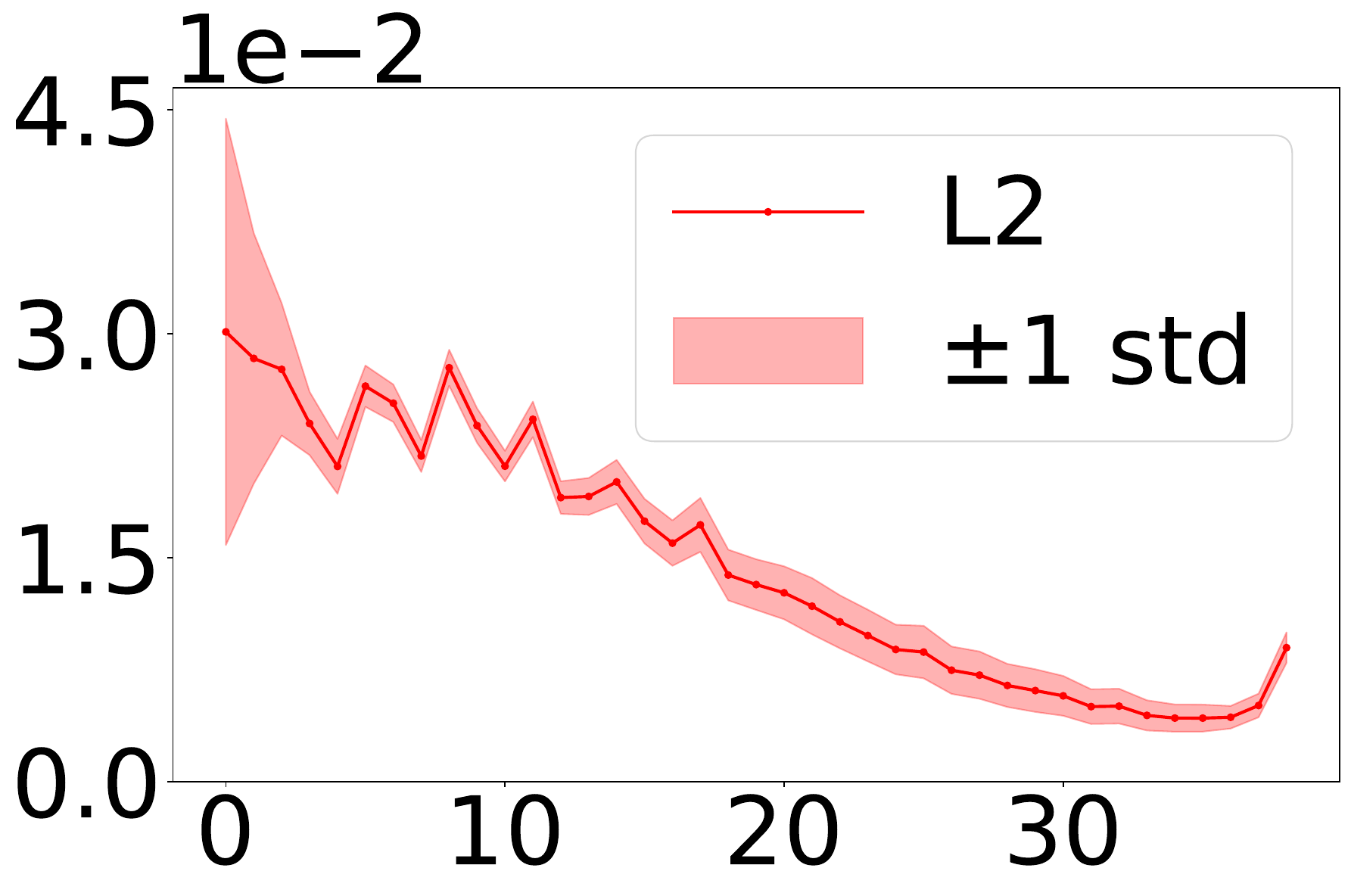}
    \caption{BG reference attention.}
    \label{U32_spatial_bg_similarity}
  \end{subfigure}
  \begin{subfigure}{0.19\linewidth}
    \includegraphics[width=\linewidth]{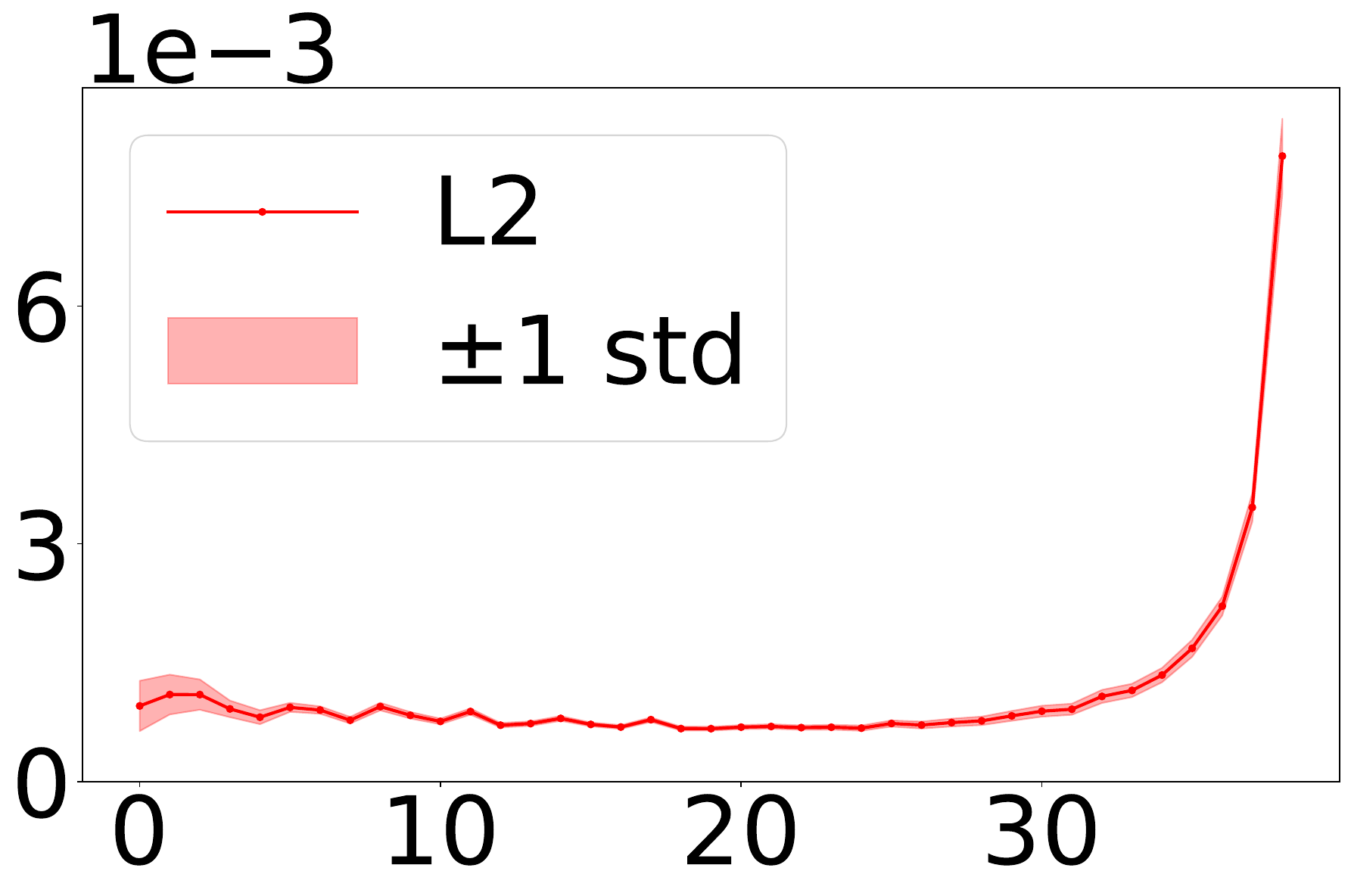}
    \caption{BG audio attention.}
    \label{U32_audio_bg_similarity}
  \end{subfigure}
  \begin{subfigure}{0.19\linewidth}
    \includegraphics[width=\linewidth]{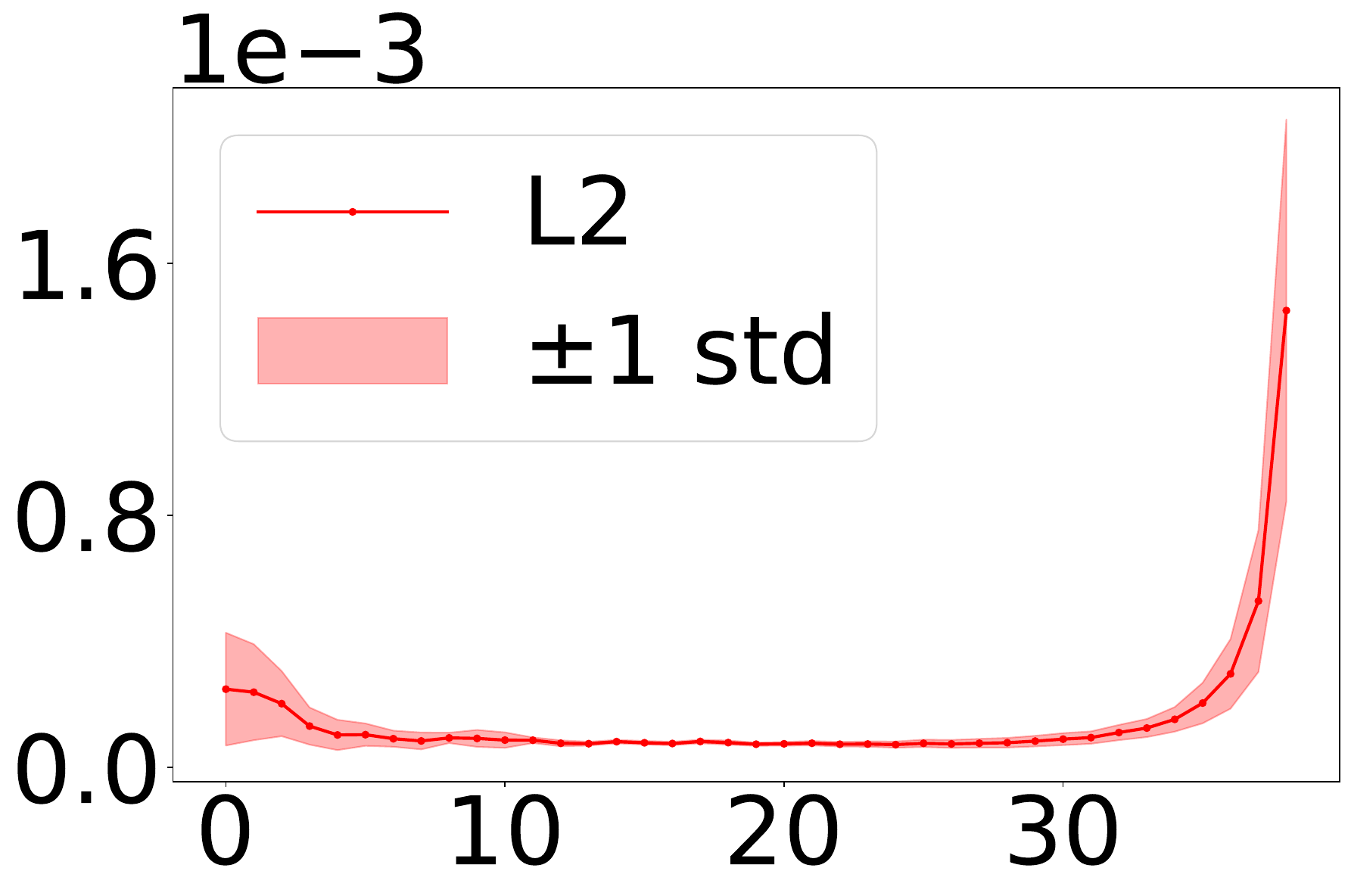}
    \caption{BG temporal attention.}
    \label{U32_temporal_1_bg_similarity}
  \end{subfigure}
  \caption{
  Average attention scores for foreground noisy latent tokens in the reference attention module ($U_{32}$) of Hallo, showing attention correlations to (a) FG and BG noisy latent features and (b) FG and BG reference features.
  The $L_2$ distance of the BG attention output features in the upsampling layer $U_{32}$ between consecutive timesteps: (c) reference attention, (d) audio attention, and (e) temporal attention. 
  \textbf{FG}: foreground. \textbf{BG}: background. \textbf{std}: standard deviation. 
  }
  \label{fig:U32_bg_attention_similarity}
\end{figure*}

More recently, DeepCache \cite{ma2024deepcache} proposes a training-free method that caches and reuses intermediate features to minimize redundant computations, achieving notable acceleration.
FasterDiffusion \cite{li2023faster} reuses encoder features to parallelize decoding across multiple timesteps. 
However, these methods do not consider the unique redundancies in talking head generation, limiting their effectiveness in this task.
To this end, we observed several critical redundancies present uniquely in talking head generation and propose a task-specific acceleration framework for maximal efficiency improvement. 
Fig. \ref{fig:feature_temporal_analysis} illustrates high temporal redundancy in the input feature $f_{U_{31}}$ to the final upsampling layer $U_{32}$, as measured by both $L_2$ distance and cosine similarity, particularly at intermediate sampling stages.
To exploit this, we propose a caching mechanism termed \textbf{Lightning-fast Caching-based Parallel denoising prediction (LightningCP)} that significantly improves inference speed without sacrificing quality. 
LightningCP caches and reuses $f_{U_{31}}$ across consecutive timesteps to bypass the entire encoder and most of the decoder, reducing model inference to merely a single upsampling layer.
We further propose to parallelize the denoising predictions of these consecutive timesteps, significantly improving inference throughput.

Moreover, talking head videos consist of foreground portrait and background environment components, which are associated with unique attention patterns. 
Fig. \ref{fig:fg_attention_weight_noise} and \ref{fig:fg_attention_weight_reference} show that foreground tokens are highly correlated with each other while exhibiting low correlation with background tokens;
In addition, Fig. \ref{U32_spatial_bg_similarity}, \ref{U32_audio_bg_similarity}, and \ref{U32_temporal_1_bg_similarity} demonstrate that the background components of the attention output features remain stable across many timesteps. 
Based on these insights, we propose \textbf{Decoupled Foreground Attention (DFA)} to perform attention exclusively with foreground tokens while reusing cached background output features, quadratically reducing the attention complexity with respect to foreground token ratio while maintaining generation quality. 

In summary, we make the following contributions:

\begin{itemize}

\item We introduce a training-free acceleration framework for talking head generation, maximizing inference speed while maintaining or even improving generation quality. 
\item We propose Lightning-fast Caching-based Parallel denoising prediction (LightningCP), a caching mechanism that reduces UNet inference to a parallelizable upsampling layer by reusing temporally stable decoder features.
\item We present Decoupled Foreground Attention (DFA) that reduces the computation complexity of attentions quadratically by leveraging the localization of foreground attention and the stability of background features.
\item Extensive experiments demonstrate that our method outperforms existing caching-based acceleration methods in terms of both efficiency and quality. 
\end{itemize}

%% file: related_works.tex
\section{Related Works}

\subsection{Diffusion-based Talking Head Generation}

Diffusion-based talking head generation can be classified into two-stage and end-to-end methods. 
Two-stage methods \cite{ma2023dreamtalk, liu2024anitalker, li2025ditto} predict a sequence of intermediate motion representations (e.g. 3DMM, implicit keypoints) from audio, and then generate talking videos from the predicted motion sequence using pretrained GAN-based generators. 
Although these methods can achieve real-time inference in some cases \cite{li2025ditto}, their generation quality is limited by the granularity of the motion representation and the rendering quality of the generator.  
In contrast, end-to-end methods \cite{tian2024emo, xu2024hallo, chen2025echomimic, zheng2024memo} leverage the powerful generative priors of latent diffusion model \cite{rombach2022high} and extend it with 3D convolutions, audio attentions, and temporal modules to generate talking head videos. 
While these methods can achieve higher generation quality, their slow inference speed hinders practical deployment. 
To bridge this gap, we propose a training-free approach for end-to-end models that yields substantial speedups while preserving quality, taking a significant step forward in real-world application.

\subsection{Diffusion Model Acceleration}


Diffusion models can be accelerated via efficient samplers \cite{song2020denoising, lu2022dpm}, model pruning \cite{fang2023structural}, and distillation \cite{salimans2022progressive, meng2023distillation}. 
However, efficient samplers do not exploit feature redundancy within diffusion models, while pruning and distillation incur expensive model training.


To address these limitations, training-free methods have gained popularity in recent years.
ToMe \cite{bolya2023token} accelerates self-attention by removing less informative tokens, but cannot accelerate talking head generation significantly due to the more complex pipeline. 
DeepCache \cite{ma2024deepcache} caches high-level features and efficiently updates only low-level features, but requires inefficient iterative denoising. 
FasterDiffusion \cite{li2023faster} enables parallel decoding for multiple timesteps by reusing encoder features, but involves costly full decoder inference. 
In contrast, our cache reduces inference to only a single decoder layer and enables parallel denoising of multiple non-key timesteps with a novel input latents estimation technique.

\subsection{Foreground-Background Decoupling}

The decoupling of foreground (FG) and background (BG) has been used in various downstream tasks of diffusion models.
FG-BG segmentation masks are used to preserve subject identity in consistent generation \cite{wang2025characonsist} and multi-prompt generation \cite{cai2025ditctrl}, as well as to preserve background in image and video editing \cite{zhu2025kv, shuai2024survey, sun2024diffusion}. 
In talking head generation, our method leverages the static nature of talking head background and employs masks to accelerate model inference for the first time. 

%% file: preliminaries.tex
\section{Preliminaries}


\subsection{Latent Diffusion Models}

Latent diffusion models (LDMs) \cite{rombach2022high} have become the foundation for high-fidelity image and video generation.
The diffusion process consists of two main stages: a forward process and a reverse process, both performed in the latent space. 
The forward process gradually adds Gaussian noise to a clean latent sample $z_0$, resulting in a sequence of increasingly noisy latents $\{z_t\}_{t=1}^T$:

$$
z_t = \sqrt{\bar{\alpha}_t} z_0 + \sqrt{1 - \bar{\alpha}_t}\, \epsilon, \qquad \epsilon \sim \mathcal{N}(0, I)
$$


\noindent Here, $z_t$ denotes the noisy latent at timestep $t$, and $\epsilon$ is standard Gaussian noise. Let $\{\beta_t\}_{t=1}^T$ be a predefined noise schedule with $\beta_t \in (0, 1)$, the signal scaling coefficient $\alpha_t$ is defined as $\alpha_t = 1 - \beta_t$, and the cumulative product $\bar{\alpha}_t$ is expressed as $\bar{\alpha}_t = \prod_{s=1}^t \alpha_s$. 

The reverse process, parameterized by a neural network, aims to reconstruct the original latent $z_0$ from $z_t$ by iteratively removing noise. 
The DDIM sampling strategy is widely adopted for efficient and deterministic inference, with the reverse update at each timestep as:



\begin{equation}
\begin{aligned}
z_{t-1} &= \lambda_t z_t +\tau_t \cdot \epsilon_\theta(z_t, t, c), \ \ \  \lambda_{t} = \sqrt{\frac{\alpha_{t-1}}{\alpha_t}}, \\ \ \tau_{t} &= \sqrt{\alpha_{t-1}} \left( \sqrt{\frac{1}{\alpha_{t-1}} - 1} - \sqrt{\frac{1}{\alpha_t} - 1} \right)
\end{aligned}
\label{eq:denoising_equation_parallel}
\end{equation}

where $\epsilon_\theta(z_t, t, c)$ is the predicted noise at timestep $t$, conditioned on $z_t$, the timestep $t$, and the condition vector $c$ that is typically a text prompt embedding.

\subsection{Diffusion-based Talking Head Generation Model}

Diffusion-based talking head generation models combine a backbone based on Stable Diffusion \cite{rombach2022high} with additional modules to perform audio-driven animation.
The backbone's denoising UNet iteratively predicts and removes noise from the noisy latent guided by the reference feature and the audio feature. 
The reference feature is extracted by Reference Net \cite{xu2024magicanimate} from the input image to ensure identity consistency throughout denoising.
To incorporate the reference feature, a reference attention module extends the self-attention of the denoising Unet by concatenating the noisy latent with the reference feature as key and value inputs.
An audio attention module injects motion information into the denoising process by performing cross-attention between the noisy latent and audio feature, producing lip movements and head motions synced with the driving audio.
A temporal layer performs self-attention on the frame dimension, capturing subtle transitions and enhancing frame-to-frame coherence. 
Finally, the denoised latent features are decoded by the VAE decoder into the talking head video frames.

At each timestep $t$ of the denoising process, the latent representation $z_t \in \mathcal{R}^{b, c, f, h, w}$ of a video clip is progressively refined, where $(b, c, f, h, w)$ represent batch size, channels, number of frames, image height, and image width. This iterative update can be formally expressed as follows:

\begin{align}
z_{t-1} &= \lambda_t z_{t} + \tau_t \cdot \epsilon_\theta(z_t, t, c), \ \ c = \{R, A, c_{\text{others}}\}
\label{eq:denoising_equation}
\end{align}

\noindent 
Specifically, the condition feature $c=\{R, A, c_{\text{others}}\}$ includes reference feature $R$, audio features $A$, and other customized condition features $c_{\text{others}}$ which may include face identity embedding \cite{xu2024hallo} or emotion features \cite{zheng2024memo}. 
In the denoising process, 
the spatial attention modules concatenate the reference feature $R$ directly with noisy latent.
Audio features $A$ spans the $f$ frames within the video clip. 
Finally, the noisy latent is reshaped from $b \times c \times f \times h \times w$ to $(b \times h \times w) \times f \times c$ as the input to the temporal layer. 
A key distinction between talking head generation and general video generation is that the former produces videos of varying lengths determined by the input audio clip duration. 
To accommodate this, the driving audio clip is segmented into fixed-length intervals, each corresponding to individual video clips of length $f$, which are then combined to form the complete video output.

%% file: methods.tex
\begin{figure}[ht]
  \centering
  \includegraphics[width=0.99\columnwidth]{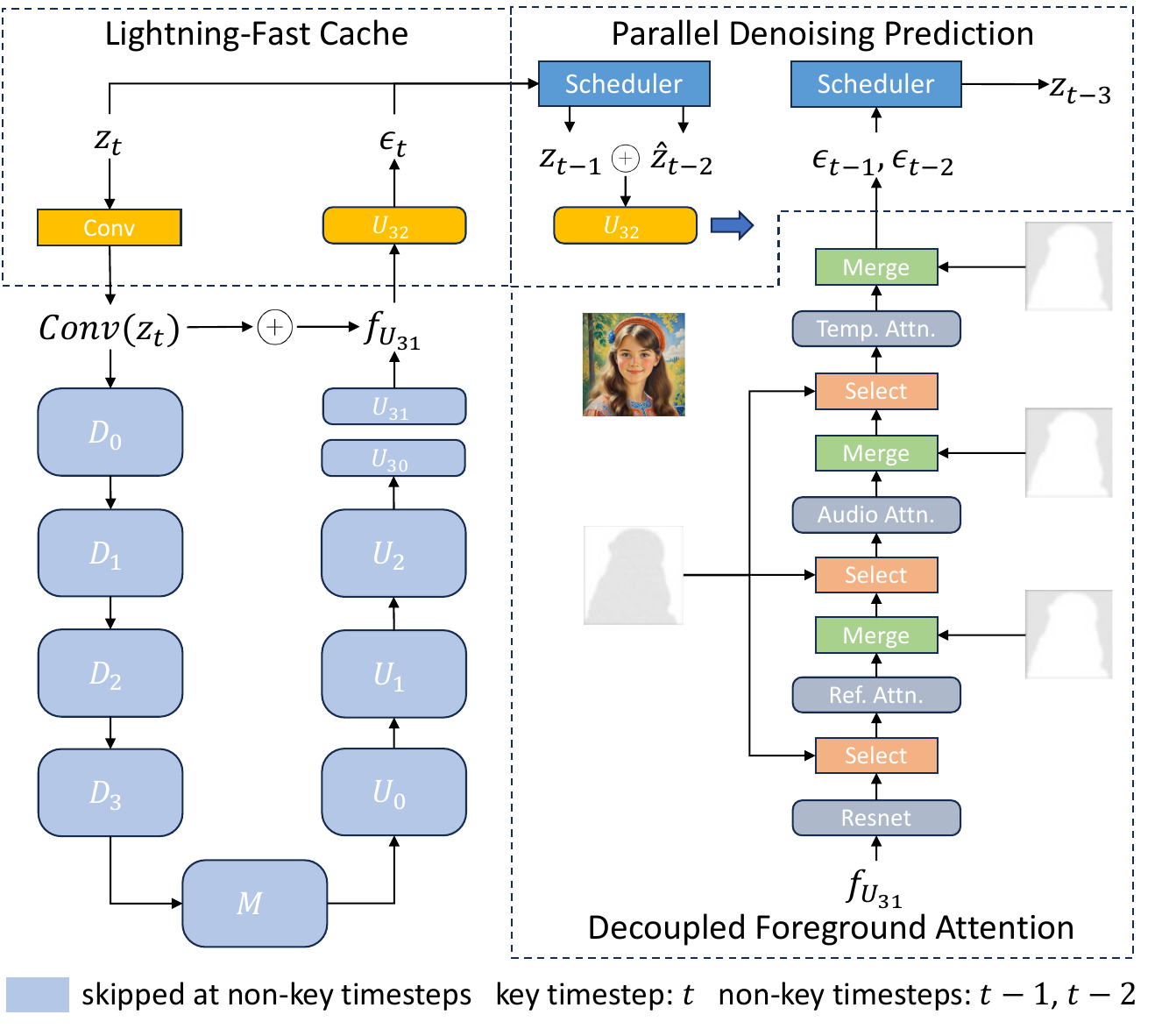}
  \caption{%
    The pipeline of the accelerated talking head model. 
    At key timestep $t$, we perform full model inference and cache feature $f_{U_{31}}$.
    At non-key timesteps $t-1$ and $t-2$, we reuse cached $f_{U_{31}}$ and bypass the encoder ($D_{0}, D_{1}, D_{2}, D_{3}$), midblock $M$, and all of the decoder ($U_{0}, U_{1}, U_{2}, U_{3}$) except its last layer $U_{32}$. 
    Moreover, denoising prediction at $t-1$ and $t-2$ can be executed in parallel and further accelerated through decoupled foreground attention. 
  }
  \label{lcp_pipeline}
\end{figure}

\section{Method}
\label{Method}

In this section, we detail the proposed acceleration methods for diffusion-based talking head generation. 
Our approach introduces Lightning-fast Caching-based Parallel denoising prediction (LCP) and Decoupled Foreground Attention (DFA), significantly reducing inference times while maintaining or even improving video quality.
In addition, auxiliary techniques, including input latent estimation and reference feature removal, are proposed to further enhance the visual quality of specific diffusion-based talking head models. 

\subsection{Lightning-fast Caching-based Parallel Denoising Prediction}
\label{lcp}


As shown in Fig. \ref{lcp_pipeline}, diffusion-based talking head generation models typically adopt a UNet architecture, with four encoder (downsampling) blocks $D_{0}, D_{1}, D_{2}, D_{3}$, a mid-block $M$, and four decoder (upsampling) blocks $U_{0}, U_{1}, U_{2}, U_{3}$.
In each denoising step, the input latent is passed sequentially through the encoder, mid-block, and decoder for denoising prediction. 
Since this full forward pass is repeated for every diffusion timestep, the process incurs substantial computation cost, especially with large latent diffusion models.


We observe that for most denoising timesteps in talking head generation, all downsampling operations in the encoder and the majority of the upsampling layers in the decoder are redundant. 
By caching and reusing high-level decoder features across timesteps, it is feasible to bypass the encoder entirely and all but the final upsampling layer of the decoder.
As illustrated in Fig. \ref{fig:feature_temporal_analysis}, the input feature $f_{U_{31}}$ to the final layer $U_{32}$ of the last upsampling block exhibits substantial temporal similarity across consecutive denoising steps. 
Specifically, Fig. \ref{f_U31_mse} depicts the feature $L_2$ distance between adjacent timesteps, while Fig. \ref{f_U31_cosine_similarity} presents the feature cosine similarity matrix computed across all timesteps. 
Both metrics confirm that $f_{U_{31}}$ maintains high temporal similarity, especially in the middle timesteps.
This insight motivates us to cache $f_{U_{31}}$ at selected key timesteps and reuse it in subsequent non-key timesteps, significantly reducing computation cost without compromising the fidelity of the generated output.

\subsubsection{Lightning-fast Cache}
\label{lightning_fast_Cache}

Motivated by this observation, we propose a lightweight caching strategy termed Lightning-fast Cache, which enables highly efficient inference by partially bypassing the U-Net architecture. 
Fig. \ref{lcp_pipeline} shows the pipeline of our proposed method.
At each key timestep $t$, we execute a full forward pass of the UNet and cache the intermediate feature $f_{U_{31}}^t$. 
Then, for the following non-key timesteps $t-1$ and $t-2$, we skip all encoder and early decoder computations by reusing the cached $f_{U_{31}}^t$.

At non-key timesteps like $t-1$, the denoising network $\epsilon_\theta$ simplifies to a subnetwork $\hat{\epsilon}_\theta$, computing only the final upsampling operation $U_{32}$ using the cached $f_{U_{31}}^t$ and a lightweight convolution of the current latent $z_t$:
\begin{align}
\epsilon_{t-1} &= \hat{\epsilon}_\theta(f_{U_{31}}^t, z_{t-1}, t-1, c) \\
&= U_{32}(f_{U_{31}}^t, \text{conv}(z_{t-1}), t-1, c)
\end{align}
where $\text{conv}(z_t)$ denotes a shallow convolution applied to the latent $z_t$ to generate the skip-connection input for $U_{32}$, and $c$ denotes additional conditioning (e.g. audio feature, reference feature). 
\subsubsection{Parallel Denoising Prediction}

\begin{figure}[h]
  \centering
  \begin{subfigure}{0.49\columnwidth}
    \includegraphics[width=\linewidth]{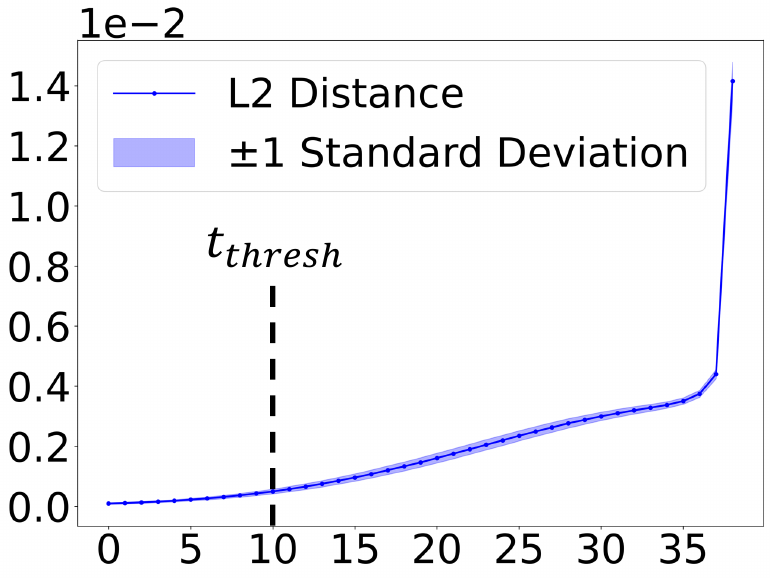}
    \caption{Input latents.}
    \label{input_latents_mse}
  \end{subfigure}
  \hfill
  \begin{subfigure}{0.49\columnwidth}
    \includegraphics[width=\linewidth]{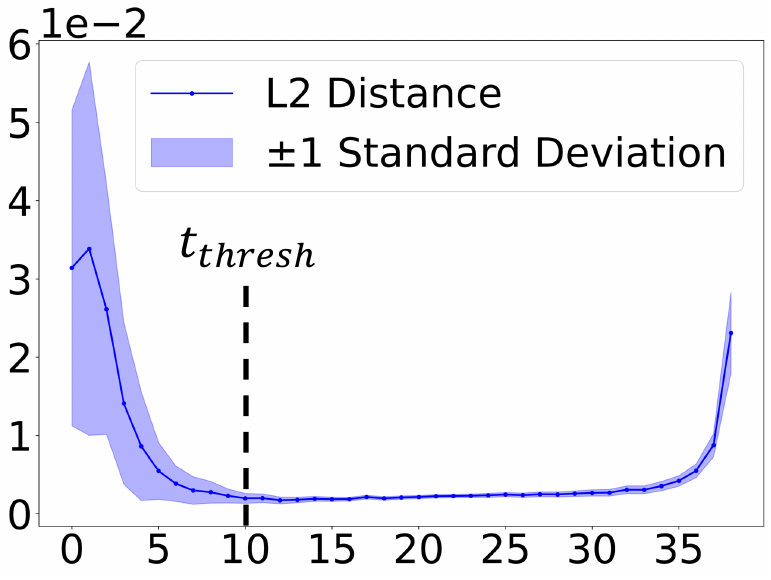}
    \caption{Predicted noise.}
    \label{noise_pred_mse}
  \end{subfigure}
  \caption{
  $L_2$ distance between consecutive timesteps for: (a) input latents, (b) predicted noise.  
  $t_{thresh}$ is the threshold timestep after which input latents estimation is applied.
  }
  \label{fig:input_latents_estimation}
\end{figure}

\label{parallel_noise_prediction}

With Lightning-fast Cache, we further accelerate denoising prediction of consecutive non-key timesteps in parallel on multiple GPUs, because inputs of these non-key timesteps share the same cached feature and differ only in their timestep-specific variables. 

Given a key timestep $t$, we cache the intermediate feature $f_{U_{31}}^t$ and reuse it across a block of consecutive non-key timesteps $\{t-1, t-2, \cdots, t-N\}$. Then, the denosing equation of non-key timestep $t-i$ becomes:

\begin{equation}
\begin{aligned}
z_{t-1-i} &= \lambda_{t-i} z_{t-i}
+ \tau_{t-i} \cdot \hat{\epsilon}_\theta(f_{U_{31}}^t, z_{t-i}, t-i, c), \\
&\text{s.t.} \ \ \ 1 \le i \le N-1. \\
\end{aligned}
\label{eq:denoising_equation_parallel}
\end{equation}

However, there is a key challenge in parallelization: when $i>1$, the input latents $z_{t-i}$ for each non-key timestep are not available in advance. 
A straightforward workaround is to approximate all unknown input latents using the latent $z_{t-1}$ computed at the key timestep.

Unfortunately, reusing the same input latent $z_{t-1}$ across multiple timesteps ${t-2, ..., t-N}$ can lead to significant error accumulation, ultimately degrading video quality. 
As illustrated in Fig. \ref{input_latents_mse}, the feature similarity between consecutive input latents decreases substantially over time when using samplers like DDIM \cite{song2020denoising}. 
This compounding discrepancy highlights the risk of not updating the input latents during parallel denoising prediction.

To address this issue, we propose \textbf{Input Latents Estimation}, which estimates input latents for consecutive non-key timesteps in parallel prediction by leveraging the temporal feature similarity of the predicted noise.
As illustrated in Fig. \ref{noise_pred_mse}, the predicted noise remain stable across the timesteps where the input latents show increasing discrepancies.
Specifically, we use the diffusion scheduler to estimate the input latents $\hat{z}_{t-2}, ..., \hat{z}_{t-N}$ for non-key timesteps $\{t-2, ..., t-N\}$ using the predicted noise $\epsilon_\theta(z_{t}, t, c)$ at key timestep $t$ through Eq. \ref{ILE_step1}, and then use them to compute the actual latents $\{z_{t-2}, ..., z_{t-N}\}$ via Eq. \ref{ILE_step2}.
For timestep $t-2$, $\hat{z}_{t-2}$ is estimated using the latent $z_{t-1}$ computed at key timestep $t$; For subsequent timesteps after $t-2$, we iteratively estimate input latents $\hat{z}_{t-i}$ using the previously estimated latent $\hat{z}_{t-i+1}$ as the reference. 



\begin{subequations}
\begin{align}
\hat{z}_{t-i} &= 
\begin{cases}
    \lambda_{t-1} z_{t-1} 
+ \tau_{t-1} \cdot \epsilon_\theta(z_{t}, t, c) & \text{for } i = 2 \\\\
    \lambda_{t-i+1} \hat{z}_{t-i+1}
+ \tau_{t-i+1} \cdot \epsilon_\theta(z_{t}, t, c) & \text{for } i > 2
\end{cases}\label{ILE_step1}\\[1ex]
z_{t-1-i} &= \lambda_{t-i} z_{t-i}
+ \tau_{t-i} \cdot \hat{\epsilon}_\theta(f_{U_{31}}^t, \hat{z}_{t-i}, t-i, c), \label{ILE_step2}\\
&\text{s.t.} \ \ \ 1 \le i \le N-1. \nonumber
\end{align}
\end{subequations}


This two-step refinement allows us to recover higher-quality output latents by using more accurate model inputs. 
In practice, we apply Input Latents Estimation after a designated threshold timestep $t_{\text{thresh}}$. In the early denoising phase ($t < t_{\text{thresh}}$), the input latent $z_t$ exhibits high similarity across steps, so parallel prediction without correction suffices. Beyond this threshold, the quality benefit of estimating $\hat{z}_{t-i}$ becomes significant. Our experiments confirm that Input Latents Estimation greatly improves video fidelity, particularly for high-speed samplers such as DDIM \cite{song2020denoising}, while incurring negligible extra computation.

\subsection{Decoupled Foreground Attention}
\label{decoupled_foreground_attention}

To further accelerate inference, we propose Decoupled Foreground Attention (DFA) to streamline the reference attention, self-attention in audio module, and temporal attentions in $U_{32}$.
The computation cost of these attentions scales quadratically with the sequence length of the video tokens, which is as large as $64 \times 64 = 4096$ in $U_{32}$.
To reduce this cost, DFA restricts the attention computation to the foreground components of the input features. 
This reduction of spatial tokens decrease both query and key length, effectively reducing the attention complexity from $\mathcal{O}(L^2)$ to $\mathcal{O}(L_f^2)$, where $L_f$ is the number of foreground tokens.


Our design is based on two empirical observations:
\begin{itemize}
    \item \textbf{Foreground-Localized Attention:} Foreground attention maps are primarily concentrated within the foreground region, indicating spatial decoupling between dynamic facial motion and static background content.
    Fig. \ref{fig:fg_attention_weight_noise} and \ref{fig:fg_attention_weight_reference} visualizes this phenomenon in the reference attention module of the upsampling layer $U_{32}$. Across all the foreground noisy latent tokens, we calculate the average attention weight sum per token with regard to tokens of 4 distinct groups: foreground noisy latent, background noisy latent, foreground reference feature, and background reference feature. The resulting analysis shows strong localization of attention within foreground tokens and weak correlation between foreground and background tokens. 
    \item \textbf{Temporal Redundancy in Background:} Background attention features remain stable across consecutive timesteps, making them suitable for caching and reuse.
    Fig. \ref{fig:U32_bg_attention_similarity} visualizes the $L_2$ distance for the background component of the attention output features in the upsampling layer $U_{32}$. 
    The analysis on reference (Fig. \ref{U32_spatial_bg_similarity}), audio (Fig. \ref{U32_audio_bg_similarity}), and temporal attentions 
    (Fig. \ref{U32_temporal_1_bg_similarity}) collectively reveal high feature similarity, particularly in the middle denoising timesteps, supporting the strategy of caching background features to reduce computational redundancy.
\end{itemize}

\paragraph{Foreground Masking for Token Reduction.}

We use a $64\times64$ face segmentation mask to separate the foreground tokens. 
The mask is downsampled from the reference image's face segmentation mask, which is obtained using an off-the-shelf face parsing model \cite{face-parsing}. 
Let $M \in \{0,1\}^{H \times W}$ denote the binary foreground segmentation mask, where an element value of $1$ indicates that the corresponding spatial location belongs to the foreground region. Given flattened attention features $Q, K, V \in \mathbb{R}^{L \times d}$, where $L = H \cdot W$, we select the foreground token subset as $Q_\mathcal{F}, K_\mathcal{F}, V_\mathcal{F} \in \mathbb{R}^{L_f \times d}$, where $L_f = \| M \|_1 = \sum_{ij} M_{ij}$.

The attention output is then computed over the reduced foreground region:
\begin{align}
A_\mathcal{F} = \text{softmax}\left( \frac{Q_\mathcal{F} K_\mathcal{F}^\top}{\sqrt{d}} \right) V_\mathcal{F} \in \mathbb{R}^{L_f \times d}.
\end{align}

\noindent This leads to a quadratic reduction in attention complexity—from $\mathcal{O}(L^2)$ to $\mathcal{O}(L_f^2)$—proportional to the foreground token ratio. In reference attention, we apply the same foreground mask to both the noise and reference features. Similarly, in the audio and temporal attentions, the computation is restricted to foreground tokens.

\paragraph{Merging with Cached Background.}
Once the updated foreground attention output $A_\mathcal{F}$ is obtained, we reconstruct the complete attention feature map by merging it with the cached background output $A_\mathcal{B}^{(t^*)} \in \mathbb{R}^{L_b \times d}$ from the most recent key timestep $t^*$, where $L_b = L - L_f$ denotes the number of background tokens.
The final combined feature is given by:
\begin{align}
A = \text{Merge}(A_\mathcal{F}, A_\mathcal{B}^{(t^*)}) \in \mathbb{R}^{L \times d},
\end{align}

By narrowing the expensive attentions to the foreground and reusing temporally stable background features, DFA achieves further acceleration with negligible impact on generation quality.


\subsubsection{Reference Feature Removal}
\label{Reference_Feature_Removal}


Reference attention incurs the highest computation cost in the entire pipeline because the contatenation of reference feature makes the key and value twice as long.
Our empirical study shows that reference features in certain layers do not affect video quality, an effect that varies among different talking head models.
In our experiments, we apply reference feature removal on all of the tested talking head models to achieve additional acceleration, while maintaining or even improving the video quality.


%% file: experiments.tex
\section{Experiments}

\begin{figure*}[h]
  \centering
  \includegraphics[width=\linewidth]{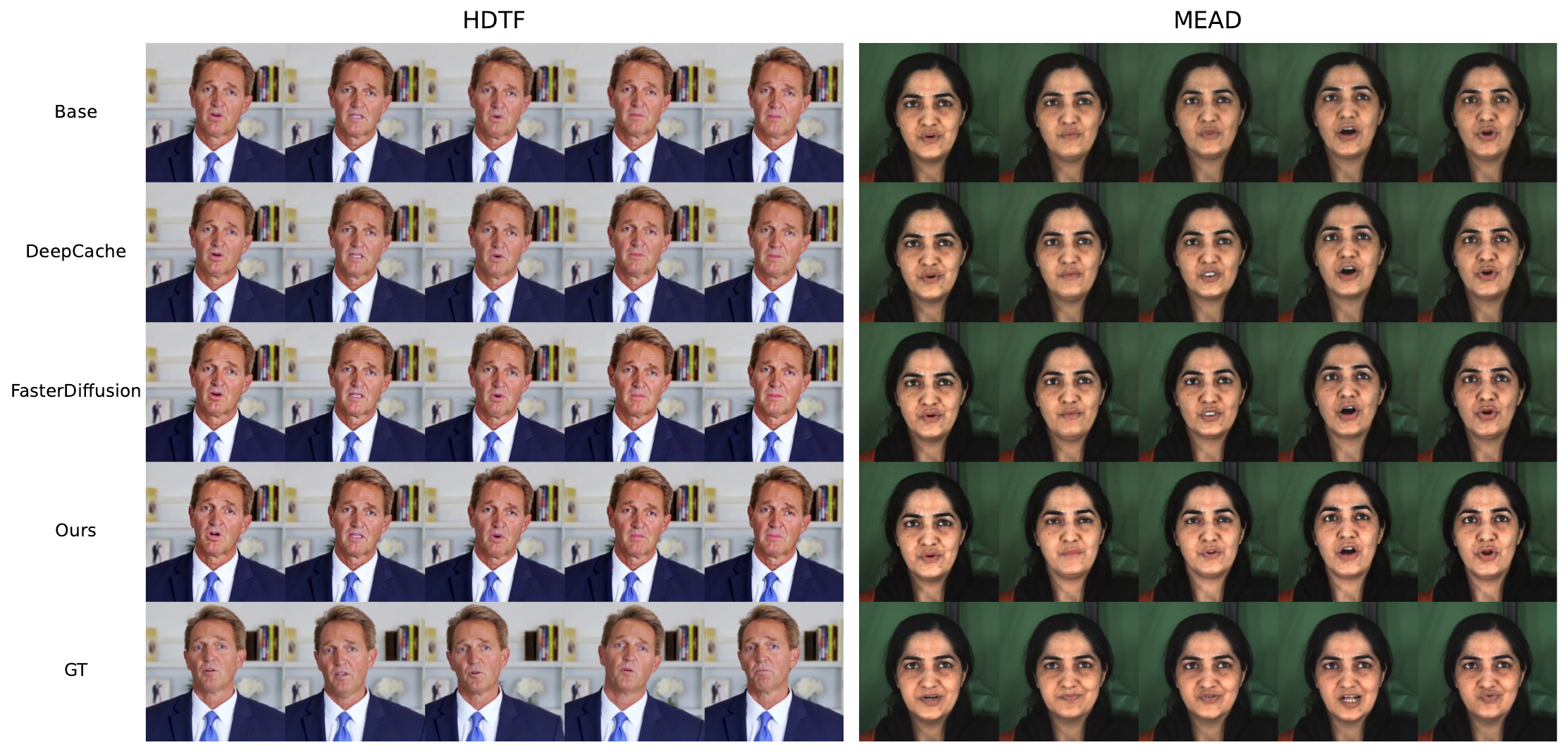}
  \caption{Qualitative result on HDTF and MEAD datasets using the Hallo model. From top to bottom: results from the base model, DeepCache, FasterDiffusion, our proposed method, and Ground Truth (GT). 
  }
  \label{fig:demo_hallo}
\end{figure*}

\input{main_result_table}

\input{ablation_result_table}

\subsection{Experimental Settings}

\paragraph{Models, Datasets, and Evaluation Metrics}

We evaluate our method on four diffusion-based talking head generation models: Hallo \cite{xu2024hallo}, MEMO \cite{zheng2024memo}, EchoMimic, and Accelerated EchoMimic (EchoMimic Acc.) \cite{chen2025echomimic}, which accelerates inference by reducing the number of diffusion steps and removing classifier-free guidance \cite{ho2022classifier}. 
For testing, we use 100 videos from the HDTF dataset \cite{zhang2021flow} and 100 videos from the MEAD dataset \cite{wang2020mead}. 

To assess video quality, we use standard evaluation metrics for talking head generation: Fréchet Video Distance (FVD) \cite{unterthiner2018towards}, Fréchet Inception Distance (FID) \cite{heusel2017gans}, expression-FID (E-FID) \cite{tian2024emo}, and Lip Sync \cite{prajwal2020lip}. 
Specifically, E-FID quantifies the expressiveness of facial motion by computing FID between 3DMM \cite{deng2019accurate} expression coefficients extracted from generated and ground-truth videos. 
To evaluate generation speed, we report latency as the UNet inference time required to generate a fixed-length video clip. 
Since the final video length varies with input audio duration, latency and FLOPs are measured per clip for fair comparison.
Speedup is computed as the latency of the base model divided by the latency of the accelerated model.

\paragraph{Baselines}

We compare our method with two training-free diffusion acceleration baselines: DeepCache \cite{ma2024deepcache} and FasterDiffusion \cite{li2023faster}. 
For fair comparison, we use identical key timesteps across all acceleration methods for each model, as timestep selection substantially affects efficiency and output quality. 
All experiments are run using A800 GPUs. 
We use a single GPU to test latency of the base model and DeepCache, while using multi-GPUs to test latency of FasterDiffusion and our method. 


\subsection{Performance Analysis}

Table~\ref{tab:main_result} presents results on HDTF \cite{zhang2021flow} and MEAD \cite{wang2020mead} datasets, comparing our method with base models and acceleration baselines \cite{ma2024deepcache, li2023faster} across multiple diffusion-based talking head models \cite{xu2024hallo, zheng2024memo, chen2025echomimic}. 
Our method achieves the most inference acceleration, with the highest speedup on Hallo (3.15×) \cite{xu2024hallo} and EchoMimic (3.10×) \cite{chen2025echomimic}. 
While MEMO \cite{zheng2024memo} and EchoMimic Acc. \cite{chen2025echomimic} have fewer diffusion steps—limiting the number of non-key timesteps—our method is still the fastest on every model. 
Furthermore, our approach achieves the lowest FLOPs in all settings, demonstrating superior computational efficiency via targeted optimizations: caching temporally stable decoder features and restricting high-cost attention operations to the dynamic foreground.

More importantly, this speedup is achieved without compromising visual quality.
As shown in Fig. \ref{fig:demo_hallo}, our method maintains the visual fidelity, identity consistency, and facial movement expressiveness of the base model. 
In addition, it facilitates the generation of more synchronized lip movements compared to the base model and the acceleration baselines. 
On the HDTF dataset, our method consistently yields the lowest FVD across all models and the highest Sync scores in most cases, while maintaining FID and E-FID comparable to the best-performing baselines. 
On MEAD, our approach achieves the lowest FID and the highest Sync scores in most cases. 
These results highlight the effectiveness of the proposed mechanisms: re-using cached decoder feature in Lightning-fast Cache preserves the feature's temporal coherency across timesteps, and Decoupled Foreground Attention preserves motion expressiveness. 
While other baselines may show competitive performance in isolated settings (e.g., DeepCache on Hallo or FasterDiffusion on EchoMimic Acc.), they often perform poorly on other models or datasets. 
In contrast, our method demonstrates consistent quality across all settings.
Additionally, FasterDiffusion requires 3 extra key timesteps on EchoMimic to avoid severe video quality degradation, significantly diminishing its speedup. 
In comparison, our method retains both high video quality and high speed under the original settings, further validating its robustness and generalization ability.

\subsection{Ablation Study}

To assess the contribution of each component of our method, we conduct ablations on HDTF \cite{zhang2021flow} and MEAD \cite{wang2020mead} using three variants: (1) Lightning Fast Caching-based Parallel denoising prediction (LCP), (2) LCP + Decoupled Foreground Attention (DFA), and (3) full model with LCP + DFA + reference feature removal (rm), as shown in Table~\ref{tab:ablation}. 

Adding LCP to the base model significantly accelerates inference speed while maintaining decent video quality. 
DFA reduces FLOPs and introduces additional speedup across all scenarios, notably achieving a 8.9\% FLOPs reduction on Hallo and pushes speedup above 3. 
Additionally, it retains or even improves video quality, indicating that removing background tokens allow the model to preserve essential interactions within the face region.
Reference feature removal further enhances lip sync, suggesting that eliminating redundant static appearance features in certain layers enables better modeling of fine-grained lip motion.

We also evaluate the effect of input latent estimation on a subset of HDTF. Table~\ref{tab:ablation_estimate} shows that without this estimation, severe degradation in FVD, FID, and E-FID occurs, confirming that the drastic decrease in the similarity of input latent features across timesteps leads to significant error accumulation.
Therefore, it is necessary to alleviate this error by estimating the input latents using the temporally stable predicted noise, which significantly improves generation quality with negligible computational overhead.

%% file: main_result_table.tex
\begin{table*}[ht]
  \centering
  \renewcommand{\arraystretch}{0.95} 
  \resizebox{\linewidth}{!}
  {
  \setlength{\tabcolsep}{3pt} 
  \begin{tabular}{@{}ll
      cccc  
      cccc  c  c  c   
    @{}}
    \toprule
    \multirow{2}{*}{Method} & \multirow{2}{*}{Variants} &
      \multicolumn{4}{c}{HDTF} &
      \multicolumn{4}{c}{MEAD} & 
      \multirow{2}{*}{\shortstack{FLOPS $\downarrow$ \\($\times 10^{12}$/clip)}} & 
      \multirow{2}{*}{\shortstack{Latency $\downarrow$ \\(s/clip)}} & 
      \multirow{2}{*}{Speedup $\uparrow$ } \\
    \cmidrule(lr){3-6} \cmidrule(lr){7-10}
      & & FVD $\downarrow$ & FID $\downarrow$ & E-FID $\downarrow$ & Sync $\uparrow$
        & FVD $\downarrow$ & FID $\downarrow$ & E-FID $\downarrow$ & Sync $\uparrow$ & \\
    \midrule
  
      
    \multirow{4}{*}{Hallo}
      & Base  
        & \textcolor{blue}{135.32} & \textcolor{blue}{4.69} & \textcolor{blue}{0.157} & \textcolor{blue}{7.38}
        & \textcolor{blue}{153.42} & \textcolor{blue}{6.17} & \textcolor{blue}{0.297} & \textcolor{blue}{6.03}
        & \textcolor{blue}{2158.08}  & \textcolor{blue}{23.692} & - \\
      & DeepCache
        & \underline{129.31} & \textbf{4.62} & \textbf{0.157} & \underline{7.52}
        & \textbf{137.56} & \textbf{6.26} & \textbf{0.276} & \underline{6.15}
        & \underline{1041.28} & 12.884 & 1.84 \\
      & FasterDiffusion 
        & 140.56 & 5.63 & 0.200 & 7.47
        & 168.72 & 7.33 & 0.285 & 6.14
        & 1561.64 & \underline{9.032} & \underline{2.62} \\
        
      \rowcolor{gray!15}
      & Ours
        & \textbf{121.80} & \underline{4.89} & \underline{0.178} & \textbf{7.66}
        & \underline{158.62} & \underline{6.64} & \underline{0.277} & \textbf{6.43}
        & \textbf{671.40} & \textbf{7.528} & \textbf{3.15} \\
    \midrule
    \multirow{4}{*}{MEMO}
      & Base
        &  \textcolor{blue}{82.07} & \textcolor{blue}{3.40} & \textcolor{blue}{0.159} & \textcolor{blue}{8.10}
        &  \textcolor{blue}{82.90} & \textcolor{blue}{5.63} & \textcolor{blue}{0.161} & \textcolor{blue}{6.06}
        & \textcolor{blue}{919.86} & \textcolor{blue}{14.934} & - \\
      & DeepCache
        &  \underline{82.52} & \textbf{3.42} & \textbf{0.162} & \underline{8.17}
        &  \underline{85.28} & \underline{5.66} & \textbf{0.170} & 6.09
        & \underline{491.14} & 9.212 & 1.62 \\
      & FasterDiffusion
        &  98.12 & 4.17 & 0.232 & \textbf{8.23}
        &  93.49 & 5.89 & 0.221 & \textbf{6.24}
        & 712.54 & \underline{6.740} & \underline{2.22} \\
      \rowcolor{gray!15}
      & Ours
        &  \textbf{80.23} & \underline{3.46} & \underline{0.163} & \underline{8.17}
        &  \textbf{79.18} & \textbf{5.54} & \underline{0.171} & \underline{6.12}
        & \textbf{406.08} & \textbf{6.416} & \textbf{2.33} \\

    \midrule
    \multirow{3}{*}{EchoMimic}
      & Base
        & \textcolor{blue}{115.78} & \textcolor{blue}{4.43} & \textcolor{blue}{0.116} & \textcolor{blue}{5.75}
        & \textcolor{blue}{166.82} & \textcolor{blue}{5.43} & \textcolor{blue}{0.335} & \textcolor{blue}{4.89}
        & \textcolor{blue}{965.19} & \textcolor{blue}{10.491} & - \\
      & DeepCache
        & 165.42 & \underline{4.57} & \textbf{0.093} & 5.86
        & 234.59 & 6.65 & \textbf{0.384} & \underline{4.83}
        & \underline{437.97} & 5.460 & 1.92 \\
      & FasterDiffusion\textsuperscript{\textdagger}
        & \underline{126.27} & 4.69 & 0.109  & \underline{5.90}
        &  \textbf{175.23} & \underline{5.67} & 0.467 & 4.79
        & 820.05 & \underline{4.788} & \underline{2.19} \\
      \rowcolor{gray!15}
      & Ours
        & \textbf{116.94} & \textbf{4.55} & \underline{0.103} & \textbf{6.08}
        & \underline{195.57} & \textbf{5.45} & \underline{0.391} & \textbf{4.93}
        & \textbf{286.77} & \textbf{3.387} & \textbf{3.10} \\
    \midrule
    \multirow{4}{*}{\shortstack{EchoMimic\\(Acc.)}}
      & Base
        & \textcolor{blue}{145.35} & \textcolor{blue}{5.06} & \textcolor{blue}{0.196} & \textcolor{blue}{6.27}
        & \textcolor{blue}{213.72} & \textcolor{blue}{6.27} & \textcolor{blue}{0.361} & \textcolor{blue}{5.33}
        & \textcolor{blue}{90.27} & \textcolor{blue}{1.0458} & - \\
      & DeepCache
        & 146.56 & \textbf{5.10} & \underline{0.203} & 6.40
        & 214.52 & \underline{6.15} & 0.367 & 5.49
        & \underline{67.49} & 0.8268 & 1.26 \\
      & FasterDiffusion
        & \underline{143.41} & 5.92 & 0.226 & \underline{6.52}
        & \textbf{157.62} & 6.65 & \textbf{0.331} & \underline{5.54}
        & 68.93 & \underline{0.7680} & \underline{1.36} \\
      \rowcolor{gray!15}
      & Ours
        & \textbf{136.14} & \underline{5.18} & \textbf{0.201} & \textbf{6.58}
        & \underline{192.90} & \textbf{5.82} & \underline{0.359} & \textbf{5.61}
        & \textbf{60.52} & \textbf{0.7236} & \textbf{1.45} \\
    \bottomrule
  \end{tabular}
  }
  \caption{Quantitative results on HDTF and MEAD. \textsuperscript{\textdagger} We include 3 extra key timesteps for FasterDiffusion on EchoMimic to avoid severe video quality degradation.}
  \label{tab:main_result}
\end{table*}

%% file: ablation_result_table.tex
\begin{table*}[]
  \centering
  \resizebox{\linewidth}{!}{%
  \begin{tabular}{@{}ll
      cccc  
      cccc  c  c  c 
    @{}}
    \toprule
    \multirow{2}{*}{Method} & \multirow{2}{*}{Variants} &
      \multicolumn{4}{c}{HDTF} &
      \multicolumn{4}{c}{MEAD} & 
      \multirow{2}{*}{\shortstack{FLOPS $\downarrow$ \\($\times 10^{12}$/clip)}} & 
      \multirow{2}{*}{\shortstack{Latency $\downarrow$ \\(s/clip)}} & 
      \multirow{2}{*}{Speedup $\uparrow$} \\
    \cmidrule(lr){3-6} \cmidrule(lr){7-10}
      & & FVD $\downarrow$ & FID $\downarrow$ & E-FID $\downarrow$ & Sync $\uparrow$
        & FVD $\downarrow$ & FID $\downarrow$ & E-FID $\downarrow$ & Sync $\uparrow$ & \\
    \midrule
    \multirow{3}{*}{Hallo}
      & LCP 
        & 124.62 & 4.99 & 0.189 & 7.53
        & 151.8 & 6.54 & 0.268 & 6.17
        & 745.72 & 8.052 & 2.94  \\
      & LCP + DFA
        & 128.31 & 5.10 & 0.198 & 7.57
        & 154.48 & 6.79 & 0.274 & 6.19
        & 679.56 & 7.688 & 3.08 \\
      & LCP + DFA + rm
        & 121.80 & 4.89 & 0.178 & 7.66
        & 158.62 & 6.64 & 0.277 & 6.43
        & 671.44 & 7.528 & 3.15 \\
    \midrule
    \multirow{3}{*}{MEMO}
      & LCP
        &  83.32 & 3.45 & 0.169 & 8.21
        &  80.26 & 5.55 & 0.168 & 6.17
        & 411.02 & 6.640 & 2.25 \\
      & LCP + DFA
        &  80.49 & 3.43 & 0.162 & 8.16
        &  79.57 & 5.50 & 0.168 & 6.11
        & 406.82 & 6.500 & 2.30 \\
      & LCP + DFA + rm
        &  80.23 & 3.46 & 0.163 & 8.17
        &  79.18 & 5.54 & 0.171 & 6.12
        & 406.08 & 6.416 & 2.33 \\
    \midrule
    \multirow{3}{*}{EchoMimic}
      & LCP
        & 136.87 & 4.50 & 0.098 & 5.96
        & 218.06 & 5.86 & 0.384 & 4.84
        & 319.26 & 3.447 & 3.04 \\
      & LCP + DFA
        & 116.33 & 4.64 & 0.099 & 5.98
        & 180.79 & 5.56 & 0.388 & 4.94
        & 291.21 & 3.405 & 3.08\\
      & LCP + DFA + rm
        & 116.94 & 4.55 & 0.103 & 6.08
        & 195.57 & 5.45 & 0.391 & 4.93
        & 286.77 & 3.387 & 3.10 \\
    \midrule
    \multirow{3}{*}{\shortstack{EchoMimic\\(Acc.)}}
      & LCP
        & 144.09 & 5.15 & 0.207 & 6.50
        & 201.47 & 5.99 & 0.363 & 5.50
        & 62.69 & 0.7368 & 1.42 \\
      & LCP + DFA
        & 138.24 & 5.20 & 0.206 & 6.47
        & 179.33 & 5.94 & 0.359 & 5.53
        & 61.33 & 0.7326 & 1.43 \\
      & LCP + DFA + rm
        & 136.14 & 5.18 & 0.201 & 6.58
        & 192.90 & 5.82 & 0.359 & 5.61
        & 60.52 & 0.7236 & 1.45 \\
    \bottomrule
  \end{tabular}}
  \caption{Ablation study results on HDTF and MEAD. \textbf{LCP}: Lightning-fast Caching-based Parallel denoising prediction. \textbf{DFA}: Decoupled Foreground Attention. \textbf{rm}: reference feature removal.}
  \label{tab:ablation}
\end{table*}

\begin{table}[ht]
  \centering
  \resizebox{\columnwidth}{!}{%
      \begin{tabular}{@{}ll
          rrrr  
        @{}}  
        \toprule
        \multirow{2}{*}{Method} & \multirow{2}{*}{Variants} &
          \multicolumn{4}{c}{HDTF} \\
        \cmidrule(lr){3-6}
          & & FVD $\downarrow$ & FID $\downarrow$ & E-FID $\downarrow$ & Sync $\uparrow$ \\
        \midrule
        \multirow{2}{*}{Hallo}
          & LCP w/o estimate
            & 208.46 & 7.14 & 0.355 & 7.64 \\
          & LCP w/ estimate
            & 200.33 & 7.04 & 0.352 & 7.63 \\
    
        \midrule
        \multirow{2}{*}{EchoMimic}
          & LCP w/o estimate
            & 460.21 & 6.66 & 0.226 & 5.98 \\
          & LCP w/ estimate
            & 225.16 & 6.22 & 0.233 & 6.05 \\
    
        \bottomrule
      \end{tabular}
  }
  \caption{Ablation study results of input latents estimation on HDTF. \textbf{w/ estimate}: with input latents estimation. \textbf{w/o estimate}: without input latents estimation.}
  \label{tab:ablation_estimate}
\end{table}

%% file: conclusion.tex
\section{Conclusion}

In this paper, we introduce Lightning-fast Caching-based Parallel denoising prediction (LightningCP), a training-free acceleration framework for diffusion-based talking head generation. 
By caching and reusing high-level decoder features, our approach enables efficient parallel inference across multiple denoising timesteps.
Additionally, Decoupled Foreground Attention (DFA) further enhances inference efficiency by restricting attention to the foreground region. 
We also employ auxiliary techniques to improve generation quality via input latents estimation and introduce extra speedup via reference feature removal.
Extensive experiment validates the superior efficiency and quality of our method, which provides a practical and plug-and-play solution for accelerating diffusion-based talking head generation.


%% file: aaai2026.bib
@inproceedings{zhang2023sadtalker,
  title={SadTalker: Learning Realistic 3D Motion Coefficients for Stylized Audio-Driven Single Image Talking Face Animation},
  author={Zhang, Wenxuan and Cun, Xiaodong and Wang, Xuan and Zhang, Yong and Shen, Xi and Guo, Yu and Shan, Ying and Wang, Fei},
  booktitle={Proceedings of the IEEE/CVF Conference on Computer Vision and Pattern Recognition},
  pages={8652--8661},
  year={2023}
}

@article{wang2021audio2head,
  title={Audio2head: Audio-driven one-shot talking-head generation with natural head motion},
  author={Wang, Suzhen and Li, Lincheng and Ding, Yu and Fan, Changjie and Yu, Xin},
  journal={arXiv},
  year={2021}
}

@inproceedings{wang2022one,
  title={One-shot talking face generation from single-speaker audio-visual correlation learning},
  author={Wang, Suzhen and Li, Lincheng and Ding, Yu and Yu, Xin},
  booktitle={Proceedings of the AAAI Conference on Artificial Intelligence},
  volume={36},
  number={3},
  pages={2531--2539},
  year={2022}
}

@inproceedings{prajwal2020lip,
  title={A lip sync expert is all you need for speech to lip generation in the wild},
  author={Prajwal, KR and Mukhopadhyay, Rudrabha and Namboodiri, Vinay P and Jawahar, CV},
  booktitle={Proceedings of the 28th ACM international conference on multimedia},
  pages={484--492},
  year={2020}
}

@inproceedings{ma2024deepcache,
  title={Deepcache: Accelerating diffusion models for free},
  author={Ma, Xinyin and Fang, Gongfan and Wang, Xinchao},
  booktitle={Proceedings of the IEEE/CVF conference on computer vision and pattern recognition},
  pages={15762--15772},
  year={2024}
}

@article{li2023faster,
  title={Faster diffusion: Rethinking the role of unet encoder in diffusion models},
  author={Li, Senmao and Hu, Taihang and Khan, Fahad Shahbaz and Li, Linxuan and Yang, Shiqi and Wang, Yaxing and Cheng, Ming-Ming and Yang, Jian},
  journal={CoRR},
  year={2023}
}

@article{salimans2022progressive,
  title={Progressive distillation for fast sampling of diffusion models},
  author={Salimans, Tim and Ho, Jonathan},
  journal={arXiv preprint arXiv:2202.00512},
  year={2022}
}

@inproceedings{meng2023distillation,
  title={On distillation of guided diffusion models},
  author={Meng, Chenlin and Rombach, Robin and Gao, Ruiqi and Kingma, Diederik and Ermon, Stefano and Ho, Jonathan and Salimans, Tim},
  booktitle={Proceedings of the IEEE/CVF Conference on Computer Vision and Pattern Recognition},
  pages={14297--14306},
  year={2023}
}

@inproceedings{fang2023structural,
  title={Structural Pruning for Diffusion Models},
  author={Fang, Gongfan and Ma, Xinyin and Wang, Xinchao},
  booktitle={Advances in Neural Information Processing Systems},
  year={2023},
  note={arXiv:2305.10924},
  url={https://arxiv.org/abs/2305.10924}
}

@inproceedings{bolya2023token,
  title={Token merging for fast stable diffusion},
  author={Bolya, Daniel and Hoffman, Judy},
  booktitle={Proceedings of the IEEE/CVF conference on computer vision and pattern recognition},
  pages={4599--4603},
  year={2023}
}

@inproceedings{tian2024emo,
  title={Emo: Emote portrait alive generating expressive portrait videos with audio2video diffusion model under weak conditions},
  author={Tian, Linrui and Wang, Qi and Zhang, Bang and Bo, Liefeng},
  booktitle={European Conference on Computer Vision},
  pages={244--260},
  year={2024},
  organization={Springer}
}

@article{xu2024hallo,
  title={Hallo: Hierarchical audio-driven visual synthesis for portrait image animation},
  author={Xu, Mingwang and Li, Hui and Su, Qingkun and Shang, Hanlin and Zhang, Liwei and Liu, Ce and Wang, Jingdong and Yao, Yao and Zhu, Siyu},
  journal={arXiv preprint arXiv:2406.08801},
  year={2024}
}

@inproceedings{chen2025echomimic,
  title={Echomimic: Lifelike audio-driven portrait animations through editable landmark conditions},
  author={Chen, Zhiyuan and Cao, Jiajiong and Chen, Zhiquan and Li, Yuming and Ma, Chenguang},
  booktitle={Proceedings of the AAAI Conference on Artificial Intelligence},
  volume={39},
  number={3},
  pages={2403--2410},
  year={2025}
}

@article{zheng2024memo,
  title={MEMO: Memory-Guided Diffusion for Expressive Talking Video Generation},
  author={Zheng, Longtao and Zhang, Yifan and Guo, Hanzhong and Pan, Jiachun and Tan, Zhenxiong and Lu, Jiahao and Tang, Chuanxin and An, Bo and Yan, Shuicheng},
  journal={arXiv preprint arXiv:2412.04448},
  year={2024}
}

@inproceedings{xu2024magicanimate,
  title={Magicanimate: Temporally consistent human image animation using diffusion model},
  author={Xu, Zhongcong and Zhang, Jianfeng and Liew, Jun Hao and Yan, Hanshu and Liu, Jia-Wei and Zhang, Chenxu and Feng, Jiashi and Shou, Mike Zheng},
  booktitle={Proceedings of the IEEE/CVF Conference on Computer Vision and Pattern Recognition},
  pages={1481--1490},
  year={2024}
}

@article{song2020denoising,
  title={Denoising diffusion implicit models},
  author={Song, Jiaming and Meng, Chenlin and Ermon, Stefano},
  journal={arXiv preprint arXiv:2010.02502},
  year={2020}
}

@article{lu2022dpm,
  title={Dpm-solver: A fast ode solver for diffusion probabilistic model sampling in around 10 steps},
  author={Lu, Cheng and Zhou, Yuhao and Bao, Fan and Chen, Jianfei and Li, Chongxuan and Zhu, Jun},
  journal={Advances in Neural Information Processing Systems},
  volume={35},
  pages={5775--5787},
  year={2022}
}

@inproceedings{rombach2022high,
  title={High-resolution image synthesis with latent diffusion models},
  author={Rombach, Robin and Blattmann, Andreas and Lorenz, Dominik and Esser, Patrick and Ommer, Bj{\"o}rn},
  booktitle={Proceedings of the IEEE/CVF conference on computer vision and pattern recognition},
  pages={10684--10695},
  year={2022}
}

@article{ho2022classifier,
  title={Classifier-free diffusion guidance},
  author={Ho, Jonathan and Salimans, Tim},
  journal={arXiv preprint arXiv:2207.12598},
  year={2022}
}

@inproceedings{zhang2021flow,
  title={Flow-guided one-shot talking face generation with a high-resolution audio-visual dataset},
  author={Zhang, Zhimeng and Li, Lincheng and Ding, Yu and Fan, Changjie},
  booktitle={Proceedings of the IEEE/CVF Conference on Computer Vision and Pattern Recognition},
  pages={3661--3670},
  year={2021}
}

@inproceedings{wang2020mead,
  title={Mead: A large-scale audio-visual dataset for emotional talking-face generation},
  author={Wang, Kaisiyuan and Wu, Qianyi and Song, Linsen and Yang, Zhuoqian and Wu, Wayne and Qian, Chen and He, Ran and Qiao, Yu and Loy, Chen Change},
  booktitle={European Conference on Computer Vision},
  pages={700--717},
  year={2020},
  organization={Springer}
}

@article{unterthiner2018towards,
  title={Towards accurate generative models of video: A new metric \& challenges},
  author={Unterthiner, Thomas and Van Steenkiste, Sjoerd and Kurach, Karol and Marinier, Raphael and Michalski, Marcin and Gelly, Sylvain},
  journal={arXiv preprint arXiv:1812.01717},
  year={2018}
}

@article{heusel2017gans,
  title={Gans trained by a two time-scale update rule converge to a local nash equilibrium},
  author={Heusel, Martin and Ramsauer, Hubert and Unterthiner, Thomas and Nessler, Bernhard and Hochreiter, Sepp},
  journal={Advances in neural information processing systems},
  volume={30},
  year={2017}
}

@inproceedings{deng2019accurate,
  title={Accurate 3d face reconstruction with weakly-supervised learning: From single image to image set},
  author={Deng, Yu and Yang, Jiaolong and Xu, Sicheng and Chen, Dong and Jia, Yunde and Tong, Xin},
  booktitle={Proceedings of the IEEE/CVF conference on computer vision and pattern recognition workshops},
  pages={0--0},
  year={2019}
}

@misc{face-parsing,
  author = {Valikhujaev Yakhyokhuja},
  title = {face-parsing},
  year = {2024},
  publisher = {GitHub},
  howpublished = {\url{https://github.com/yakhyo/face-parsing}},
  note = {GitHub repository}
}

@inproceedings{li2025ditto,
  title={Ditto: Motion-space diffusion for controllable realtime talking head synthesis},
  author={Li, Tianqi and Zheng, Ruobing and Yang, Minghui and Chen, Jingdong and Yang, Ming},
  booktitle={Proceedings of the 33rd ACM International Conference on Multimedia},
  pages={9704--9713},
  year={2025}
}

@article{ma2023dreamtalk,
  title={Dreamtalk: When expressive talking head generation meets diffusion probabilistic models},
  author={Ma, Yifeng and Zhang, Shiwei and Wang, Jiayu and Wang, Xiang and Zhang, Yingya and Deng, Zhidong},
  journal={arXiv preprint arXiv:2312.09767},
  volume={2},
  number={3},
  year={2023}
}

@inproceedings{liu2024anitalker,
  title={Anitalker: animate vivid and diverse talking faces through identity-decoupled facial motion encoding},
  author={Liu, Tao and Chen, Feilong and Fan, Shuai and Du, Chenpeng and Chen, Qi and Chen, Xie and Yu, Kai},
  booktitle={Proceedings of the 32nd ACM International Conference on Multimedia},
  pages={6696--6705},
  year={2024}
}

@inproceedings{wang2025characonsist,
  title={Characonsist: Fine-grained consistent character generation},
  author={Wang, Mengyu and Ding, Henghui and Peng, Jianing and Zhao, Yao and Chen, Yunpeng and Wei, Yunchao},
  booktitle={Proceedings of the IEEE/CVF International Conference on Computer Vision},
  pages={16058--16067},
  year={2025}
}

@inproceedings{cai2025ditctrl,
  title={Ditctrl: Exploring attention control in multi-modal diffusion transformer for tuning-free multi-prompt longer video generation},
  author={Cai, Minghong and Cun, Xiaodong and Li, Xiaoyu and Liu, Wenze and Zhang, Zhaoyang and Zhang, Yong and Shan, Ying and Yue, Xiangyu},
  booktitle={Proceedings of the Computer Vision and Pattern Recognition Conference},
  pages={7763--7772},
  year={2025}
}

@article{zhu2025kv,
  title={Kv-edit: Training-free image editing for precise background preservation},
  author={Zhu, Tianrui and Zhang, Shiyi and Shao, Jiawei and Tang, Yansong},
  journal={arXiv preprint arXiv:2502.17363},
  year={2025}
}

@article{tu2025global,
  title={Global and local semantic completion learning for vision-language pre-training},
  author={Tu, Rong-Cheng and Ji, Yatai and Jiang, Jie and Kong, Weijie and Cai, Chengfei and Zhao, Wenzhe and Wang, Hongfa and Yang, Yujiu and Liu, Wei},
  journal={IEEE Transactions on Pattern Analysis and Machine Intelligence},
  year={2025},
  publisher={IEEE}
}

@article{tu2025prospective,
  title={Prospective layout-guided multi-modal online hashing},
  author={Tu, Rong-Cheng and Mao, Xian-Ling and Liu, Jin-Yu and Ma, Zi-Ao and Lan, Tian and Huang, Heyan},
  journal={IEEE Transactions on Image Processing},
  year={2025},
  publisher={IEEE}
}

@article{sun2025vorta,
  title={VORTA: Efficient Video Diffusion via Routing Sparse Attention},
  author={Sun, Wenhao and Tu, Rong-Cheng and Ding, Yifu and Jin, Zhao and Liao, Jingyi and Liu, Shunyu and Tao, Dacheng},
  journal={arXiv preprint arXiv:2505.18809},
  year={2025}
}

@article{liu2025t2v,
  title={T2V-OptJail: Discrete Prompt Optimization for Text-to-Video Jailbreak Attacks},
  author={Liu, Jiayang and Liang, Siyuan and Zhao, Shiqian and Tu, Rongcheng and Zhou, Wenbo and Liu, Aishan and Tao, Dacheng and Lam, Siew Kei},
  journal={arXiv preprint arXiv:2505.06679},
  year={2025}
}

@article{sun2024asymrnr,
  title={Asymrnr: Video diffusion transformers acceleration with asymmetric reduction and restoration},
  author={Sun, Wenhao and Tu, Rong-Cheng and Liao, Jingyi and Jin, Zhao and Tao, Dacheng},
  journal={arXiv preprint arXiv:2412.11706},
  year={2024}
}

@article{tu2024spagent,
  title={Spagent: Adaptive task decomposition and model selection for general video generation and editing},
  author={Tu, Rong-Cheng and Sun, Wenhao and Jin, Zhao and Liao, Jingyi and Huang, Jiaxing and Tao, Dacheng},
  journal={arXiv preprint arXiv:2411.18983},
  year={2024}
}

@article{shuai2024survey,
  title={A survey of multimodal-guided image editing with text-to-image diffusion models},
  author={Shuai, Xincheng and Ding, Henghui and Ma, Xingjun and Tu, Rongcheng and Jiang, Yu-Gang and Tao, Dacheng},
  journal={arXiv preprint arXiv:2406.14555},
  year={2024}
}

@article{sun2024diffusion,
  title={Diffusion model-based video editing: A survey},
  author={Sun, Wenhao and Tu, Rong-Cheng and Liao, Jingyi and Tao, Dacheng},
  journal={arXiv preprint arXiv:2407.07111},
  year={2024}
}

@inproceedings{tu2025automatic,
  title={Automatic evaluation for text-to-image generation: Task-decomposed framework, distilled training, and meta-evaluation benchmark},
  author={Tu, Rong-Cheng and Ma, Zi-Ao and Lan, Tian and Zhao, Yuehao and Huang, He-Yan and Mao, Xian-Ling},
  booktitle={Proceedings of the 63rd Annual Meeting of the Association for Computational Linguistics (Volume 1: Long Papers)},
  pages={22340--22361},
  year={2025}
}
